# The substitution effects on electronic structure of iron selenide superconductors

A. Ciechan[a,∗], M.J. Winiarski[b], M. Samsel-Czekała[b]

[a]*Institute of Physics, Polish Academy of Sciences, al. Lotników 32/46, 02-668 Warsaw, Poland*
[b]*Institute of Low Temperature and Structure Research, Polish Academy of Sciences, ul. Okólna 2, 50-422 Wroclaw, Poland*

**Abstract**

The influence of a partial substitution with S, Te, Co, Ni and Cu atoms on the electronic structure of the FeSe superconductor has been investigated within the density functional theory. The results of the supercell calculations reveal distinct changes of electronic structures of the substituted FeSe systems, which can be responsible for their superconducting properties. The replacement of Se atoms by Te or S ones yields imperfect nesting between the holelike and electronlike Fermi surface (FS) sheets, which enhances magnetic fluctuations responsible for superconducting pairing, thus leading to higher values of the superconducting critical temperatures. Meanwhile, the substitutions with transition-metal atoms for iron sites make more substantial changes of the FSs topology, since the holelike cylinders shrink at the cost of an enlargement of the electronlike ones. Thus, the superconducting pairing, driven by the nesting between these sheets, weakens and superconductivity disappears for a small percentage of dopants. The results support the idea of spin-fluctuation mediated superconductivity in iron chalcogenides.

## 1. Introduction

The discovery in 2008 y. of superconductivity (SC) in fluorine-doped LaFeAsO [1] attracted a lot of attention and rich families of iron-based superconductors have been detected. Among them, FeSe iron chalcogenides [2] seem to be a promising series of superconductors. The latter containing no arsenic atoms, unlike pnictide superconductors, are particularly important for applications. They are also convenient for theoretical investigations because of simple both chemical compositions and crystal structures.

The pure FeSe system is superconducting below the critical temperature $T_c$ = 8 K [2]. In its solid solutions with tellurium, FeSe$_{1-x}$Te$_x$, the $T_c$ is raised up to 15 K for $x$ =0.5 [3, 4, 5]. The end member, FeTe ($x$ =1), is no longer superconducting, but shows an antiferromagnetic phase at low temperatures. Additionally, $T_c$ increases to 37 K for FeSe [6, 7, 8, 9, 10] and to 26 K for FeSe$_{0.5}$Te$_{0.5}$ [11, 12] under external pressure of 9 GPa and 2 GPa, respectively. The sulphur substitution into FeSe up to 20% also slightly enhances $T_c$ (~ 10 K) [5], while its substitution into FeTe compound induces SC with $T_c$ ~ 8 K after suppressing the spin ordering [13, 14].

∗Corresponding author. Tel.: +48 22 1163 191; fax: +48 22 8430 926.
*Email addresses:* `ciechan@ifpan.edu.pl` (A. Ciechan), `M.Winiarski@int.pan.wroc.pl` (M.J. Winiarski), `M.Samsel@int.pan.wroc.pl` (M. Samsel-Czekała)

It turned out that the properties of SC in these chalcogenides may be tuned by some disorder [15, 16, 17], caused by e.g. the excess iron atoms in Fe$_{1+x}$Se (deficiency of selenium in FeSe$_{1-x}$) layers [18, 19]. Note that SC exists in nearly stoichiometric FeSe. The doping with 5-10% of magnetic Ni and Co ions on Fe sites in polycrystalline FeSe [5, 20] and with even less percentage in FeSe$_{1-x}$Te$_x$ single crystals [21] completely suppresses SC. A similar effect is observed for doping with non-magnetic Cu atoms for fractions above 1.5-3% [14, 21, 22, 23]. Nonetheless, superconductivity in a Cu-doped FeSe alloy is restored by external pressure and reaches its maximum above 30 K at 7.8 GPa [24].

Both photoemission spectroscopy [25, 26, 27, 28] and theoretical studies of chalcogenides [29, 30, 31, 32] have indicated an existence of a few bands in the vicinity of the Fermi level, $E_F$, which originate mainly from the Fe 3$d$ states. Since the superconducting gaps open on the electronlike as well as holelike Fermi surface (FS) sheets [27, 28], the multigap nature of SC in FeSe is associated with its multiband structure.

It was postulated that in these Fe-based superconductors, the imperfect FS nesting between its electronlike and holelike cylindrical sheets may be responsible for superconducting pairing, mediated by antiferromagnetic spin fluctuations, being connected with the **q** ~ ($\pi, \pi$) wave vector [29, 31, 32, 33]. The FS changes induced by disorder [34, 35, 36] particularly influence such nesting-driven interband interactions.

The main aim of this paper is to examine the electronic structure of several substituted FeSe systems in the normal state by the density functional theory (DFT) calculations in the supercell and rigid band approaches. We discuss structural parameters obtained by a geometry optimization for the pure FeSe and its alloys. We are especially interested in the changes of the FS topology, which is suspected to determine spin fluctuations and, thus, also be correlated with the $T_c$'s of these iron chalcogenides.

## 2. Computational details and structural properties

Band structure calculations for FeSe superconductors have been performed in the framework of DFT in the local-density approximation (LDA) of the exchange-correlation potential [37]. We used the pseudopotential method, based on plane-waves and Projector-Augmented Waves (PAW), implemented in the QUANTUM ESPRESSO code [38]. Total energy of the investigated systems was converged with accuracy to $10^{-4}$ Ry for 50 Ry energy cut-off of the plane-wave basis. The self-consistent field (SCF) and density of states (DOS) calculations were performed with the 16×16×16 **k**-point mesh in the non-equivalent part of the Brilouin zone (BZ), whereas the Fermi surface cuts were obtained with much denser **k**-point grids up to 201 × 201 in *ab*-planes of the BZ. Prior to computations of all electronic properties, both lattice parameters and atomic positions in the unit cell (u.c.) and supercells were optimized with 0.05 GPa convergence criterion on the pressure and $10^{-3}$ Ry/Bohr on forces.

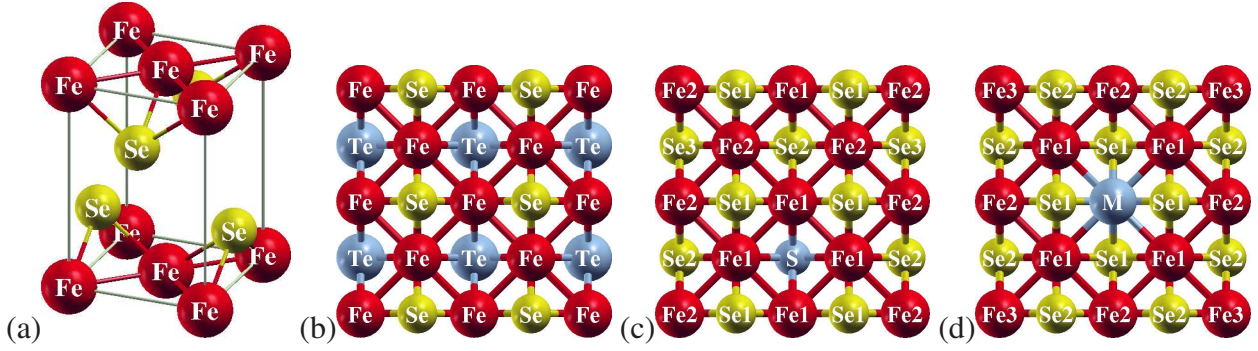

Figure 1: (a) The unit cell of tetragonal FeSe of PbO-type (space group *P*4/*nmm*). The *xy*-projection of 2×2×1 supercells used for: (b) FeSe$_{0.5}$Te$_{0.5}$, (c) FeSe$_{0.875}$S$_{0.125}$ and (d) Fe$_{0.875}$M$_{0.125}$Se with $M$ = Co, Ni or Cu atom. Non-equivalent Se and Fe positions around doping atoms are indicated in part (c) and (d).

The unit cell of pure FeSe in the tetragonal phase of the PbO-type (space group *P*4/*nmm*) is displayed in Fig. 1(a). The Fe ions form a square lattice and the Se ions are located either directly above or below the Fe plane, thus the u.c. contains two Fe and two Se atoms, i.e. double formula unit (f.u.). The results of full geometry relaxations for FeSe are given in table 1. The lattice paramters *a* and *c* are shorter than experimental ones [2, 5, 8], which is typical of the LDA usage. However, the obtained atomic position $z_{Se}$ is much better reproduced compared with the results of calculations limited to optimization of the free $z_{Se}$ parameters [25, 29, 34, 39]. The importance of the chalcogen-distance from the iron plane (and hence $z_{Se}$) has been indicated in both experimental [10] and theoretical [40] works.

In the case of partially substituted systems, we used the $2 \times 2 \times 1$ supercell (Fig. 1(b)-(d)) to simulate 12.5 atomic % of Ni, Co, Cu and S as well as 50 atomic % of Te contents.

For tellurium-doped selenides, we investigate the FeSe$_{0.5}$Te$_{0.5}$ composition, which exhibits maximum $T_c$ in the whole FeSe$_{1-x}$Te$_x$ series [3, 4, 5]. The local atomic structure study via the pair density function (PDF) analysis of neutron diffraction data [41] and the extended x-ray absorption fine-structure EXAFS measurements [42] indicate that the Se and Te ions do not share the same atomic site, which leads to different $z_{Se}$ and $z_{Te}$ coordinates. Hence, the local symmetry of FeSe$_{1-x}$Te$_x$ is lower than the average *P*4/*nmm*. It is consistent with our results yielding the Te-distance from the iron plane, $z_{Te}$, longer than $z_{Se}$ as given in Table 1. Since the experimental data had revealed the Te atoms having been randomly distributed in the crystal [41], in our calculations, we tested different arrangements of Te ions in the supercell. As the computed total energies for these cases were comparable, we used further only the structure with the alternate arrangement of the considered atoms (Fig. 1(b)). Additionally, this arrangement requires only a replacement of one Se by Te atom in the single u.c. and might be realized in both the $1 \times 1 \times 1$ u.c. and $2 \times 2 \times 1$ supercell. The crystals with other atomic configurations have much lower symmetries, which is not convenient for our study of nesting properties.

Similarly to the Te - substitution, the sulphur atoms are also incorporated into the Se-sublattice. The $T_c$ for 10% and 20% of S content is slightly enhanced from 8 K to about 8.6 and 9.9 K, respectively [5]. Thus, we realize here 12.5% substitution, which corresponds to one S atom in the supercell (Fig. 1(c)).

Table 1: Calculated and experimental lattice parameters $a$, $c$, and free $z_{Se}$ ($z_{Te/S}$ in parenthesis) atomic positions in unit cells of pure FeSe and its alloys. In the case of systems with S or $M$ = Co, Ni, Cu atoms, the mean values of selenium position, $<z_{Se}>$, are shown.

| System | $a$[Å] | $c$[Å] | $z_{Se}$ |
|---|---|---|---|
| FeSe | 3.596 | 5.431 | 0.256 |
| ref. [2] | 3.765 | 5.518 | |
| ref. [5] | 3.7696 | 5.520 | |
| ref. [8] | 3.766 | 5.499 | 0.266 |
| FeSe$_{0.5}$Te$_{0.5}$ | 3.655 | 5.685 | 0.238 (0.289) |
| ref. [21] | 3.799 | 6.056 | |
| ref. [41] | 3.800 | 5.954 | 0.244 (0.285) |
| FeSe$_{0.875}$S$_{0.125}$ | 3.590 | 5.398 | 0.258 (0.233) |
| ref. [5] (10% S) | 3.763 | 5.503 | |
| Fe$_{0.875}$Co$_{0.125}$Se | 3.603 | 5.303 | 0.260 |
| ref. [5](10% Co) | 3.764 | 5.504 | |
| Fe$_{0.875}$Ni$_{0.125}$Se | 3.607 | 5.283 | 0.262 |
| ref. [5] (10% Ni) | 3.771 | 5.503 | |
| Fe$_{0.875}$Cu$_{0.125}$Se | 3.606 | 5.319 | 0.264 |
| ref. [23] (10% Cu) | 3.841 | 5.512 | 0.259 |

In turn, the transition-metal atoms are located at the Fe-ions sites (Fig. 1(d)) and have a negative impact on $T_c$ [5, 23]. Samples with Co substitution over 10% show no superconducting state. The 5%, and 3% concentrations of Ni and Cu atoms, respectively, also completely suppress $T_c$. The considered here $2 \times 2 \times 1$ supercell calculations allow for simulations of 12.5% dopant atoms concentration, which is sufficient for investigations of electronic structure changes in the Co-substituted systems. Any considerations of lower contents of Ni or Cu atoms require larger supercells, however, in that approach the band-folding effects lead to significant difficulties in studies of the FS nesting with a vector $\mathbf{q} \sim (\pi, \pi)$. Thus, for Ni and Cu dopped systems, we used only $2 \times 2 \times 1$ supercells to predict possible general trends related to these dopants.

Our calculated lattice parameters and free atomic positions for different solid solutions, compared with the experimental ones available in literature, are collected in Table 1. Note that most of the crystallographic data was obtained at room temperature from polycrystalline samples and the final chemical compositions can be slightly different from those presented in the references. Our calculation results show that except for the Te substitution, all lattice parameters differ only a little from the pure FeSe. This tendency is in good agreement with the x-ray diffraction data. Additionally, the optimization of supercells yields non-equivalent Se and Fe positions around dopant atoms (indicated in Fig. 1(c) and (d)), which leads to substantially lower symmetry than the initial PbO-type structure and to somewhat different Fe-Se distances. In general, $M$-Se bonds are a little longer than the average Fe-Se bonds in Fe$_{1-x}M_x$Se alloys, whereas Fe-S bonds are slightly shorter than other Fe-Se ones in FeSe$_{1-x}$S$_x$. Thus, we present here only average Se positions in the supercells, $<z_{Se}>$.

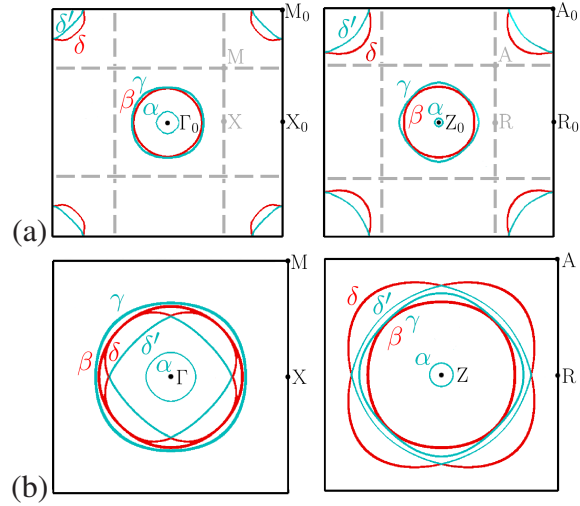

Figure 2: Fermi surface sections of FeSe through special **k**-points in (a) BZ *vs.* (b) FBZ boundaries. In part (a) the FBZ boundaries are also indicated by dashed grey line. Note that $M_0$ and $A_0$ points in BZ are folded to $\Gamma$ and Z points, respectively, in FBZ wedges. The most nested sheets $\beta$ and $\delta$ are marked by black (red) curves.

All optimized parameters are used to compute band energies, densities of states (DOS), Fermi surfaces and their nesting properties of the FeSe and its alloys in the supercell approach. Additionally, FeSe and FeSe$_{0.5}$Te$_{0.5}$ unit cells are considered for studying spin susceptibilities of the systems.

## 3. Electronic-structure results within supercell approach

### 3.1. Fermi surface of pure FeSe

In recent years, the electronic structure of the pure FeSe compound has been studied in a number of works [9, 29, 34, 39] indicating a few bands crossing the Fermi level. The electronlike and holelike cylindrical FS sheets are spanned by the nesting vector $\mathbf{q} \sim (\pi, \pi)$. Such FS nesting makes possible the antiferromagnetic spin fluctuations, being responsible for the inter-band pairing. Thus, the connection between FS topology and superconductivity is still under debate.

In Fig. 2, the FS of the pure FeSe is visualized in the form of $\Gamma_0 X_0 M_0$- and $Z_0 R_0 A_0$-plane sections in the tetragonal BZ, which corresponds to the $1 \times 1 \times 1$ u.c. and then in the $\Gamma$XM- and ZRA-plane sections in the folded Brillouin zone (FBZ) of the $2 \times 2 \times 1$ supercell. As is visible in Fig. 2(a), the FS in the BZ boundaries contains three holelike ($\alpha$, $\beta$, $\gamma$) and two electronlike ($\delta$, $\delta'$) sheets centered along the $\Gamma_0$-$Z_0$ and $M_0$-$A_0$ lines, respectively. As seen in Fig. 2(b), the sheets around the $M_0$ and $A_0$ points in the BZ are folded to the $\Gamma$ and Z points in the FBZ boundaries. Notice that now all five sheets are centered just around the $\Gamma$-Z line and the nested sheets overlap. Thus, the area of overlapping contours in the FBZ reflects an intensity of the ideal nesting, which occurs with $\mathbf{q} = (\pi, \pi)$ vector in the BZ of the $1 \times 1 \times 1$ u.c. It makes the supercell approach convenient for studying the nature of nesting.

For the pure FeSe, the nesting is clearly seen for the bands $\beta$ and $\delta$, which are marked by black (red), while the other ones - by grey (cyan). However, the nesting is strong in the $\Gamma$XM-plane and considerably weaker in some higher $k_z$-planes. In the ZRA-plane, the ideal nesting with $\mathbf{q} = (\pi, \pi)$ is not visible at all (complete lack of overlapping), which is caused by the more tree-dimensional (more corrugated) electronlike cylinder than the holelike one.

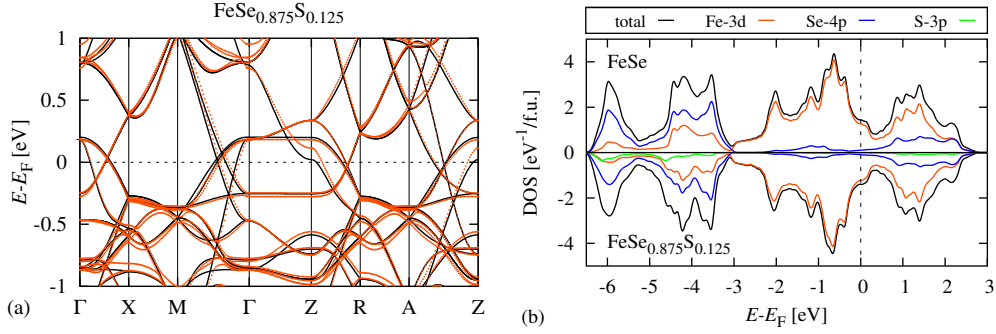

Figure 3: FeSe$_{0.875}$S$_{0.125}$: (a) band structure (pure FeSe bands are marked by solid black lines) and (b) total and averaged over non-equilibrium Fe and Se positions orbital-projected DOSs (compared with those of FeSe).

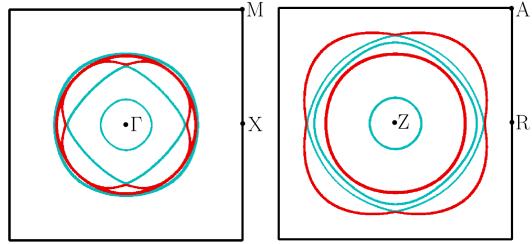

Figure 4: Fermi Surface of FeSe$_{0.875}$S$_{0.125}$ in ΓXM and ZRA planes in tetragonal FBZ boundaries.

All further results of FS calculations for partially substituted FeSe systems are presented only in the FBZ convention. In this convention, an easy distinction between strongly nested bands ($\beta$ and $\delta$) and the other ones simplifies the insight into FS changes taking place in the solid solutions.

### 3.2. Effects of substitution with sulphur and tellurium

The calculated band structure of FeSe$_{0.875}$S$_{0.125}$ is presented in Fig. 3(a). It is clear from this figure that the substitution with an S atom for Se one insignificantly changes bands in the vicinity of the Fermi level. The computed DOS at $E_F$ in pure FeSe is equal to about 1.44 states/eV/f.u. while the partial replacement of Se positions by S atoms slightly decreases this value to 1.40 states/eV/f.u. and has no impact on the overall shape of the DOS plotted in Fig. 3(b). Similarly, the FS remains almost unchanged by doping with 12.5% of S atoms - compare Fig. 4 with Fig. 2(b). Thus, the small modifications of the FS topology lead to only negligible enhancement of $T_c$, observed experimentally in [5].

Oppositely, the distinct impact of substitution is manifested in the FeSe$_{0.5}$Te$_{0.5}$ solid solution. In this system, the band splitting just around $E_F$ is presented in Fig. 5(a). Te 5$p$ electrons fill in the DOS pseudogap around 3 eV below $E_F$, which is depicted in Fig. 5(b). The total DOS at $E_F$ increases to 1.71 states/eV/f.u., which is correlated with a $T_c$ enhancement in Fe(Se,Te). In general, such behaviour might indicate the phonon mechanism of superconductivity. However, former studies on the pure electron-phonon coupling as an SC mechanism in iron-based superconductors yielded unrealistic $T_c$ less than 1 K [29, 43].

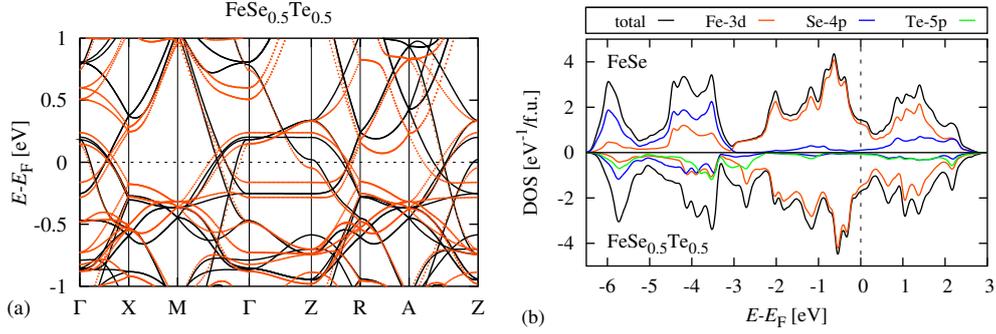

Figure 5: Same as in Fig. 3 but for FeSe$_{0.5}$Te$_{0.5}$.

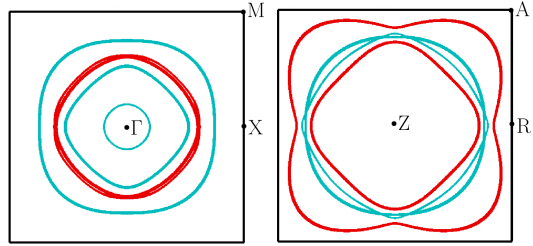

Figure 6: Same as in Fig. 4 but for FeSe$_{0.5}$Te$_{0.5}$.

In analogy to the effect of applying hydrostatic pressure on FeSe [30, 31], the FS cylinders in FeSe$_{0.5}$Te$_{0.5}$ [30, 32] are more corrugated along the $k_z$ direction, which can be connected with the changes of spin fluctuations in this compound with respect to FeSe. The overlay of the FS sheets is diminished even in the ΓXM-plane in relation to the pure FeSe. It is caused by the imperfect matching in size of electron and hole cylinderlike sheets with radius differing by a small $\delta \mathbf{q}$. However, the sections of the FS are still parallel to each other and can provide an optimal condition for nesting with $\mathbf{q} = (\pi, \pi)$ and the appearance of enhanced spin fluctuations in the system containing tellurium.

*3.3. Effects of substitution with cobalt, nickel and copper*

The band-structure results of Co, Ni and Cu substitution for Fe atoms, presented in Fig. 7, point out that the considered here transition-metal concentration of 12.5% distinctly modifies the electronic structure with respect to FeSe. It is caused by a presence of the iron-derived bands at the vicinity of $E_F$, which makes that all bands are downshifted compared with the parent-compound, proportionally to the number of additional electrons in the doped systems. As presented in Fig. 7(c), this effect is the most pronounced in the case of the copper substitution, where all hole bands around the Γ point lie well below $E_F$.

The DOS at the Fermi level in relation to FeSe is almost unchanged in Fe$_{0.875}$Co$_{0.125}$Se. Meanwhile, 12.5% of Ni and Cu atoms decreases the DOS value to 1.36 and 1.06 states/eV/f.u., respectively. The overall total DOSs of all the three considered compounds, plotted in Fig. 8(a), exhibit only minor shifts towards lower energies compared with FeSe. This stability is based on the Fe 3$d$ orbitals present in all three systems - see Fig. 8(b). In turn, the 3$d$-electron contributions from dopant atoms, shown in Fig. 8(c), are spread out at considerably lower energies than the more itinerant Fe 3$d$ electrons. In particular, almost completely filled Cu 3$d$-shells have negligible contributions to the DOS at $E_F$.

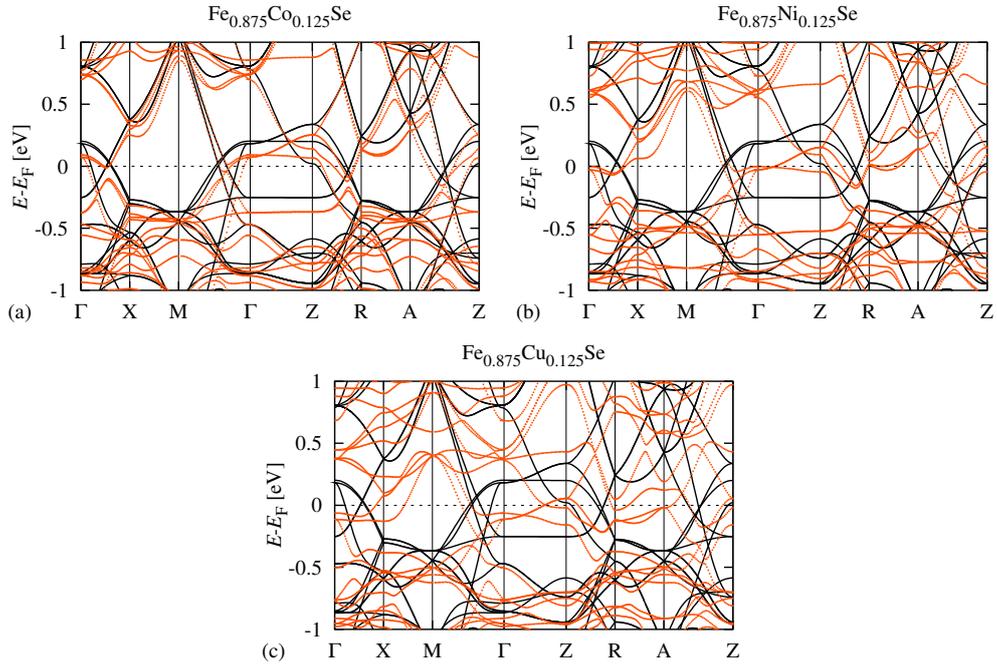

Figure 7: Band structures of (a) $Fe_{0.875}Co_{0.125}Se$, (b) $Fe_{0.875}Ni_{0.125}Se$ and (c) $Fe_{0.875}Cu_{0.125}Se$, compared with pure FeSe bands marked by solid black lines.

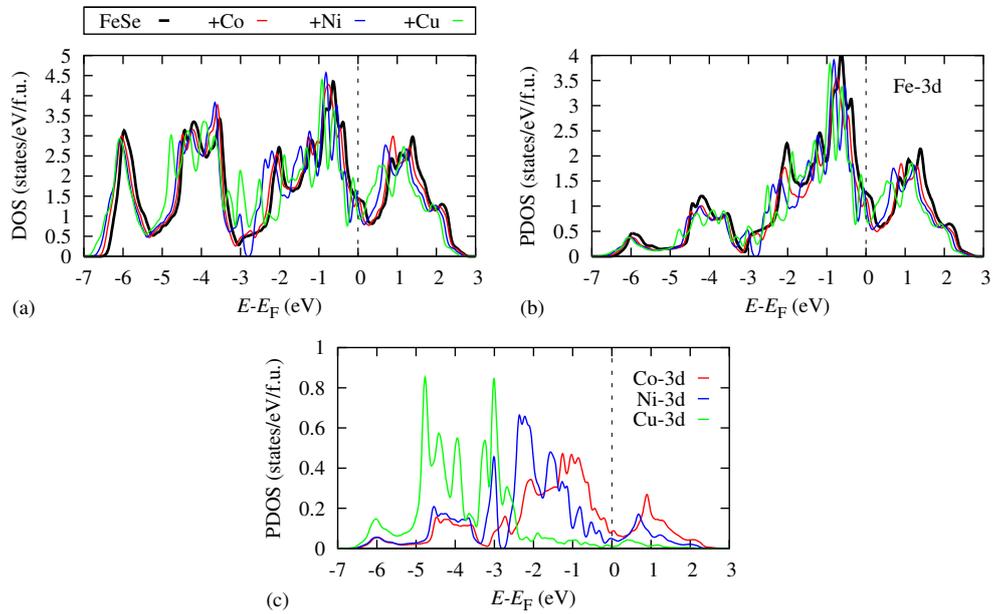

Figure 8: DOSs of FeSe doped with 12.5% of Co, Ni and Cu atoms: (a) total and partial (PDOSs) for: (b) Fe (averaged over Fe sites) and (c) dopant-atom $3d$ orbitals.

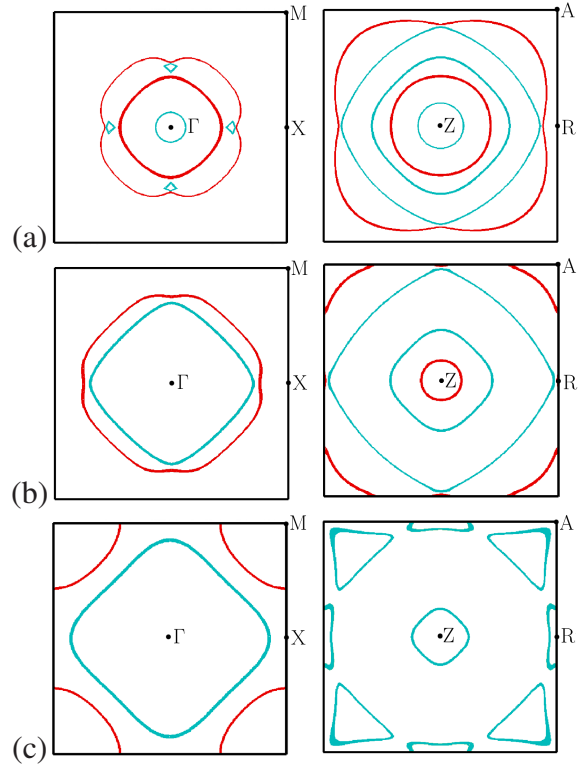

Figure 9: Fermi surface sections of 12.5% (a) Co-, (b) Ni- and (c) Cu-doped FeSe, drawn in tetragonal FBZ boundaries.

Although the studied here compositions with 12.5% content of Ni or Cu dopant elements are far above the limits of vanishing SC, they allow to detect the possible influence on the FS topology. Fig. 9 demonstrates that the main FS-sections of all alloys are substantially different from that of the pure FeSe (Fig. 2(b)).

In Fe$_{0.875}$Co$_{0.125}$Se (Fig. 9(a)), the $\beta$ and $\delta$ sheets still exist but they are now very distant from each other thereby suppressing the FS nesting. In addition, some other FS-sheets occurring in the pure FeSe are now absent in the $\Gamma$XM-plane and, hence, the FS topology has also been changed. It is a consequence of shrinking all the holelike cylinders at the cost of enlarging the electronlike ones. The modifications are even more considerable in the other systems studied in this subsection.

Namely, in Fe$_{0.875}$Ni$_{0.125}$Se (Fig. 9(b)), the holelike $\beta$ sheet around the $\Gamma$ point does not occur and only a small pocket appears around Z point. The electronlike $\delta$ sheet is, in turn, strongly enlarged and crosses the FBZ boundaries in the ZRA plane. Similar results are obtained for Fe$_{0.875}$Cu$_{0.125}$Se, where the FS topology is dramatically changed (see Fig. 9(c)), because only
9

one of the nested sheets, $\delta$, is present in the ΓXM plane and some new bands contribute to the Fermi surface. In both alloys, the nesting properties typical of the parent FeSe are completely absent.

The comparison between these systems indicates that the FSs are rapidly changed at much earlier stage when doping with Ni and Cu than with Co atoms, which destroys the possibility of the FS nesting for a smaller content of these elements in FeSe. This finding is strongly correlated with a suppression rate of $T_c$ in these compounds.

## 4. Spin susceptibility

Based on our calculation FS results for the 2×2×1 supercell, FeSe and FeSe$_{0.5}$Te$_{0.5}$ exhibit the imperfect nesting with $\mathbf{q} \sim (\pi, \pi)$ vector, spanning the holelike and electronlike FS sheets, which is usually associated with appearance of antiferromagnetic spin fluctuations. This nesting for the latter compound is reduced mainly by three-dimensionality of the FS as well as by differences in the shape and size between its cylinders, visible even in the ΓXM plane.

The nesting-driven spin fluctuations can be more precisely estimated in the single u.c. (1×1×1) by calculating the Lindhard spin susceptibility, defined by equation:

$$\chi(\mathbf{q}, \omega) = -\frac{1}{(2\pi)^3} \sum_{\mathbf{k} \in BZ} \frac{f(\varepsilon_\mathbf{k}) - f(\varepsilon_{\mathbf{k+q}})}{\varepsilon_\mathbf{k} - \varepsilon_{\mathbf{k+q}} - \omega - i\delta}. \qquad (1)$$

Figs. 10 (a) and (b) show the static form of the response function $\chi(\omega \to 0)$ at $q_z = 0$ for both FeSe and FeSe$_{0.5}$Te$_{0.5}$ compounds. A peak at M$_0$ $(\pi, \pi)$, induced by transitions from the holelike to electronlike sheets, is consistent with their FS topology. A broadening of the peak in both cases is caused by an imperfect nesting. Interestingly, FeSe$_{0.5}$Te$_{0.5}$ exhibits a peak of stronger intensity at the nesting wave vector, in spite of being a little wider. As superconducting pairing is associated with an integration of $\chi$ over the Fermi surface [29, 44], the result indicates an enhancement of the spin fluctuations and superconducting pairing in FeSe$_{0.5}$Te$_{0.5}$, being usually competitive to a spin density wave (SDW) ordering.

To consider the similar characteristics for FeSe doped with transition elements Co, Ni and Cu, a rigid-band approximation has been utilized. The similar method was successfully employed to nonstoichiometric Fe$_{1+\delta}$Te leading to reproducing a proper magnetic order in these compounds [36], being in accord with the neutron scattering measurements [45].

The strong changes in the FSs topology of the Co-doped FeSe are mainly due to the downshift of the bands compared with the parent compound (Fig. 7). Especially, the strongly nested bands $\beta$ and $\delta$ are well reproduced in the supercell calculations after the upshift of the Fermi level by about 0.1 eV. For a large doping with Ni and Cu, the changes are more pronounce, however, some band shift is still visible. In addition, a small contribution of these transition metals to DOSs around $E_F$ is negligible (particular for Cu, see Fig. 8) justifying an application of a rigid band technique, especially for interesting, smaller dopant contents. It is worth noting that the concentrations of 10% Co, 5% Ni and 3.3% Cu corresponds to doping by 0.1 electrons per formula unit and the Fermi level shift by about 0.07 eV. Taking into account small differences in the cell parameters even for 12.5% doping concentrations (see Table 1), we used the ones obtained for the pure FeSe.

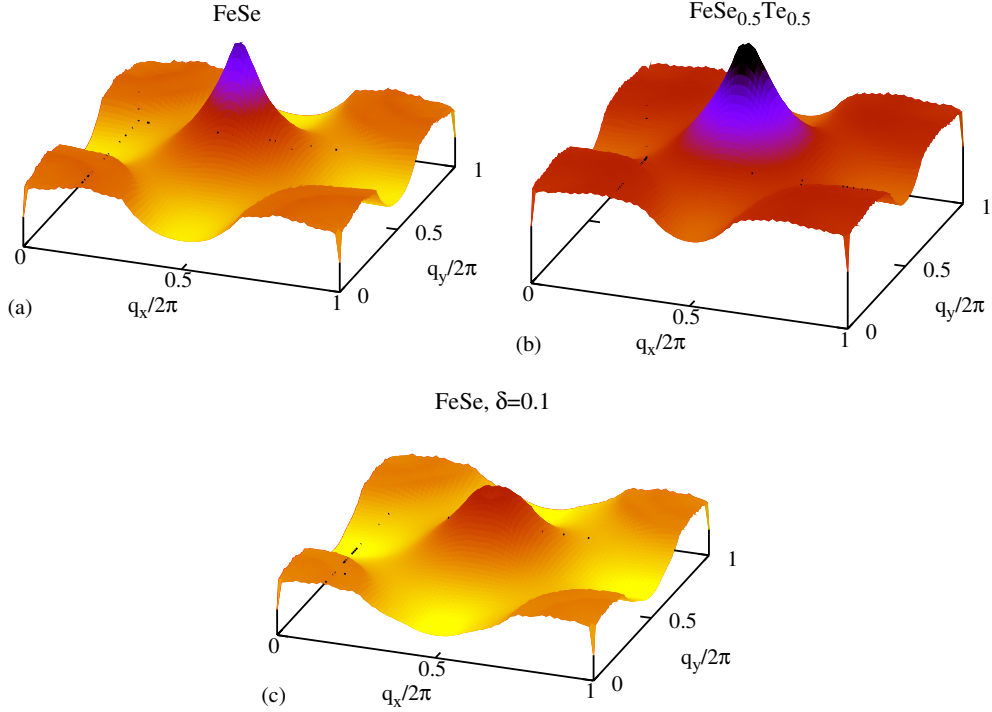

Figure 10: The real part of the noninteracting spin susceptibilities $\chi(\omega \to 0)$ (in arbitrary units) for (a) FeSe, (b) FeSe$_{0.5}$Te$_{0.5}$ and (c) FeSe with additional $\delta = 0.1$ electron/f.u.

The obtained spin susceptibility is shown in the Fig. 10(c). It is quite consistent with the supercell calculation for 12.5% Co, where the FS sheets do not overlap at all, because their size mismatch is substantial. There are still some perpendicular sections, thus, the nesting with the vector $\mathbf{q} \sim (\pi, \pi)$ does not disappear completely but is very weakened in comparison with the FeSe and FeSe$_{0.5}$Te$_{0.5}$ compounds. With further doping, the susceptibility at M$_0$ is still suppressed and becomes comparable to that at $\Gamma_0$ (0, 0), derived from intra- and inter-band interactions between two hole or two electron sections of the FS.

Thus, the nature of the nesting between the holelike and electronlike FS sheets and, hence, also the spin fluctuations are strongly suppressed with doping, which is correlated with weakening superconducting pairing and an observed suppression of $T_c$.

## 5. Conclusions

FeSe exhibits the FS nesting with the vector $\mathbf{q} \sim (\pi, \pi)$. The ideal nesting, in general, gives a strong and narrow peak in the spin susceptibility $\chi(\mathbf{q})$ and an ordered spin density wave (SDW) rather than a superconducting state is expected in such a system [29]. The fact that the nesting is reduced for FeSe seems to be crucial from the point of view of a competition between the SDW ordering and superconducting pairing. Furthermore, when the SDW order is supressed, spin fluctuations of the SDW-type can exist and mediate in the superconducting pairing.

In the paper, we have investigated effects of a substitution with S, Te atoms into Se sites as well as Co, Ni and Cu atoms into Fe sites on the electronic structure of such solid solutions. It should be pointed out that these substitutions into Fe and Se sites lead to qualitatively different effects on the electronic structure as well as superconductivity of FeSe.

The substitution of S and Te atoms for the Se sites insignificantly modifies the electronic properties of the pure FeSe compound and the superconducting properties. However, the Fermi surface sheets of FeSe$_{0.5}$Te$_{0.5}$ become more corrugated in the $k_z$ direction and the FS nesting apperas to be more imperfect. Moreover, the spin susceptibility shows a peak at M$_0$ being both broader and of a higher intensity. Therefore, a possible magnetic fluctuations and the superconducting phase in Fe(Se,Te) solid solutions can appear at an earlier stage.

Oppositely, the substitution of the transition-metal atoms (Co,Ni,Cu) for Fe sites completely modifies the Fermi surface topology and the other electronic properties, compared with the parent FeSe compound. The holelike cylinders shrink and electronlike sheets enlarge and, hence, the spin fluctuation responsible for superconducting pairing between these sheets become insufficient - superconductivity is destroyed at a small percentage of a given dopant.

The findings indicate that the electron band structure plays a crucial role in the superconductivity of the Fe-based chalcogenides and it confirms the spin fluctuation mechanism of their SC pairing.

## Acknowledgments

The authors acknowledge Prof. Piotr Bogusławski for valuable discussions. This work has been supported by the EC through the FunDMS Advanced Grant of the European Research Council (FP7 Ideas) as well as by the National Center for Science in Poland (Grant No. N N202 239540 and No. DEC-2011/01/B/ST3/02374). Calculations were partially performed on ICM supercomputers of Warsaw University (Grant No. G46-13) and in Wroclaw Centre for Networking and Supercomputing (Project No. 158). The atomic structures and Fermi Surfaces were prepared with the XCRYSDEN program [46].